\documentclass[prl,twocolumn,superscriptaddress,showpacs,floatfix,amssymb,amsmath]{revtex4}

\usepackage{times}
\usepackage{amssymb}
\usepackage{graphicx}
\usepackage{dcolumn}
\usepackage[latin1]{inputenc}
\usepackage[mathscr]{eucal}
\usepackage{epsfig}
\usepackage{rotating}
\usepackage{bm}
\usepackage{textcomp}
\usepackage{hyperref}

\begin{document}
\title{Creation of Dirac monopoles in spinor Bose-Einstein condensates}

\author{Ville Pietil\"a}
\affiliation{Department of Applied Physics/COMP, Helsinki
  University of Technology P.~O.~Box 5100, FI-02015 TKK, Finland}
\affiliation{Australian Research Council, Centre of Excellence for
Quantum Computer Technology, The University of New South Wales,
Sydney 2052, Australia}
\author{Mikko M\"ott\"onen}
\affiliation{Department of Applied Physics/COMP, Helsinki
University of Technology P.~O.~Box 5100, FI-02015 TKK, Finland}
\affiliation{Australian Research Council, Centre of Excellence for
Quantum Computer Technology, The University of New South Wales,
Sydney 2052, Australia} \affiliation{Low Temperature Laboratory,
Helsinki University of Technology, P.~O.~Box 3500, FI-02015 TKK,
Finland}

\begin{abstract}
We demonstrate theoretically that using standard external magnetic
fields, one can imprint point-like topological defects to the spin
texture of a dilute Bose-Einstein condensate. Symmetries of the
condensate order parameter render this topological defect to be
accompanied with a vortex filament corresponding to the Dirac
string of a magnetic monopole. The vorticity in the condensate
coincides with the magnetic field of a magnetic monopole,
providing an ideal analogue to the monopole studied by Dirac.
\end{abstract} 


\maketitle

The existence of particles with non-zero magnetic charge, that is,
magnetic monopoles has far-reaching implications to the laws of
quantum mechanics, theories of elementary particles, and
cosmology~\cite{Dirac:1931,Vilenkin:1994,Guth:1981}. However,
experimental evidence of magnetic monopoles as fundamental
constituents of matter is still absent, and hence there is great
incentive to search corresponding configurations in experimentally
more tractable systems. Several aspects of monopoles have
been investigated in the context of liquid
crystals~\cite{Chuang:1991}, anomalous quantum-Hall
effect~\cite{Fang:2003}, and exotic spin
systems~\cite{Castelnovo:2008}. Recently, Qi {\it et al.} proposed that a 
magnetic monopole could be induced in a topological insulator~\cite{Qi:2009}. 
Despite these elegant proposals, an experimental realization of a magnetic
monopole as an emergent particle or any analog of the Dirac monopole in still 
lacking. One of the most promising
candidate systems to realize such analogies has been superfluid
$^3$He~\cite{Blaha:1976,Volovik:1976,Salomaa:1987,Volovik:2003},
but to date there are no direct experimental observations of these
topological excitations.

Dilute Bose-Einstein condensates (BECs) of alkali atoms with a
hyperfine spin degree of freedom combine magnetic and superfluid
order and share many common features with the superfluid
$^3$He~\cite{Volovik:2003,Ohmi:1998,Ho:1998,Stenger:1998}. The order parameter 
describing such systems is typically invariant under global symmetries that 
form a non-Abelian group and monopoles and other textures can occur if this 
symmetry is broken. On the other hand, spinor BECs are well-suited to host 
artificially generated gauge fields which can provide an alternative method to 
realize a magnetic monopole~\cite{Ruseckas:2005}. In the simplest case of 
spin-$1$ condensate, a variety of different topological
defects such as global monopoles~\cite{Stoof:2001}, non-Abelian magnetic
monopoles~\cite{Pietila:2008b}, global textures~\cite{Kawaguchi:2008} and in 
particular, analogies to the Dirac monopole~\cite{Savage:2003} have been 
investigated. In the related two-component condensates, skyrmion textures have 
been studied by several 
authors~\cite{Ruostekoski:2001,AlKhawaja:2001,Battye:2002}. 
An experimental realization of any of these topological states
still remains a milestone in the field of cold atoms.

In this Letter, we consider a spinor BEC with the total hyperfine
spin $F=1$ which in the absence of external magnetic fields has
two phases: a ferromagnetic and an antiferromagnetic (polar)
phase~\cite{Ho:1998}. However, in the presence of a strong enough external 
magnetic field, the spin of the order parameter aligns with the local field 
and the condensate order parameter corresponds to the ferromagnetic phase. 
Modifying the external field adiabatically, multi-quantum vortices can be
imprinted into the condensate as was proposed theoretically by
Nakahara et al.~\cite{nakaharaetal} and verified experimentally by
Leanhardt et al.~\cite{leanhardtetal}. The method we employ here utilizes the 
same ideology in this respect, but due to non-trivial three-dimensional
structure of the magnetic field we are able to create a point-like
defect to the spin texture of the condensate that gives rise to vorticity  
coinciding with the magnetic field of a magnetic monopole.

Let us assume first that the hyperfine spin of the condensate is aligned with 
an external magnetic field which is a combination of two quadrupole fields 
and a homogeneous bias field
\begin{equation}
\label{mg-field} \bm{B}(\bm{r},t) = B_1'(x\hat{\bm{x}} +
y\hat{\bm{y}}) + B_2'z\hat{\bm{z}} + B_0(t)\hat{\bm{b}},
\end{equation}
where Maxwell's equations impose the condition $2B_1'+B_2' = 0$
for the magnetic field gradients. The direction of the bias field
is determined by the unit vector $\hat{\bm{b}}$. Such combination of 
quadrupole fields is produced by, e.g., a crossing pair of Helmholtz coils,
and it has recently been proposed to dynamically generate
knot-like textures in antiferromagnetic
BECs~\cite{Kawaguchi:2008}.
The point where the external magnetic field vanishes is the center
of a monopole defect in the spin texture
$\bm{\mathcal{S}}=\bm{\Psi}^{\dagger}\bm{\mathcal{F}}\bm{\Psi}$
given by the condensate order parameter $\bm{\Psi} =
(\psi_1,\,\psi_0,\psi_{-1})^T$ and the spin-$1$ matrices
$\bm{\mathcal{F}} =
(\mathcal{F}_x,\,\mathcal{F}_y,\,\mathcal{F}_z)$. This monopole
can be characterized by the charge~\cite{Savage:2003}
\begin{equation}
\label{charge} \mathcal{Q} =
\frac{1}{8\pi}\int_{\Sigma}\mathrm{d}^2\sigma_i\,
\varepsilon_{ijk}^{}\varepsilon_{abc}^{}\hat{s}_a\partial_j\hat{s}_b
\partial_k\hat{s}_c,
\end{equation}
where $\hat{\bm{s}} = \bm{\mathcal{S}}/|\bm{\mathcal{S}}|$ and the
integral is taken over a surface $\Sigma$ enclosing the
defect. The indices take values $1,2,3$ and summation over
repeated indices is implied. The Levi-Civita tensor is denoted by
$\varepsilon_{\alpha\beta\gamma}^{}$. For the magnetic field in
Eq.~(\ref{mg-field}), the charge of the spin texture becomes $\mathcal{Q}=1$.
A schematic illustration of
two possible textures is shown in Fig.~\ref{monopoles}.
\begin{figure}[h!]
\centering
\includegraphics[width=0.35\textwidth]{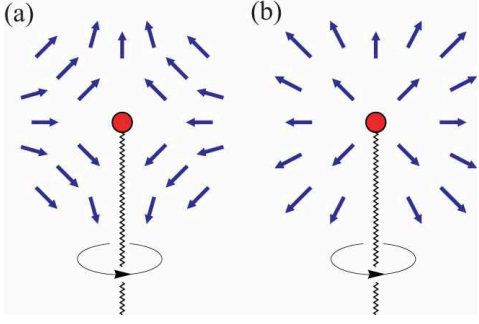}
\caption{\label{monopoles} Possible configurations for the spin
  texture. Both configurations are characterized by
the same charge $\mathcal{Q}=1$ but only the one in
(a) can be imprinted using the combination of two quadrupole
fields. The hedgehog texture in (b) describes the vorticity
$\bm{\Omega}_s$ corresponding to the spin texture in panel (a). In
both cases, the vector fields are symmetric with respect to
rotations about the axis on which the singularity filament (zigzag
line) lies.}
\end{figure}
For the ferromagnetic order parameter manifold, the second homotopy group is 
trivial and isolated monopole defects are not allowed. Thus  
the monopole defect in the spin texture is associated with a vortex filament 
extending outwards from the monopole giving rise to a physical Dirac
string~\cite{Dirac:1931,Savage:2003}. Since the monopole is
located at the zero of the magnetic field, adjusting the bias
field $B_0(t)$ moves the monopole along the axis given by
$\hat{\bm{b}}$. In particular, the monopole can be brought in from
outside of the condensate by taking large enough bias to begin
with and then ramping it down adiabatically.

The superfluid velocity can be defined as $\bm{v}_s^{} = -
i\hbar\,\zeta^{\dagger}\nabla \zeta/m$, where  $\bm{\Psi} =
\sqrt{\varrho}\,\zeta$, $\varrho$ is the density of particles, and
$m$ is the mass of the constituent
atoms~\cite{Ho:1998,Savage:2003,Ho:1996}. For simplicity, we first
assume that $\hat{\bm{b}} =  \hat{\bm{z}}$ and define new
coordinates by $(x',y',z') = (x,y,2z-B_0(t)/B_1')$. In the adiabatic limit, the 
condensate order parameter corresponds to the local eigenstate 
of the linear Zeeman operator $g_{\scriptscriptstyle F}\mu_{\scriptscriptstyle B}|\bm{B}|\,\hat{\bm{n}}\cdot\bm{\mathcal{F}}$ with  
$\hat{\bm{n}} = (\sin\beta\cos\alpha,\,\sin\beta\sin\alpha,\,\cos\beta)$. The 
local Zeeman state can be written as 
\begin{equation}
\label{eigenstate}
|\zeta(\alpha,\beta)\rangle = \mathcal{U}(\alpha,\beta)|F,m_{F}^{}\rangle,
\end{equation}
where the unitary transformation $\mathcal{U}(\alpha,\beta)$ is given by
$\mathcal{U}(\alpha,\beta) = e^{-i\alpha\mathcal{F}_z}e^{-i\beta\mathcal{F}_y}
e^{i\alpha\mathcal{F}_z}$ and the state $|F,m_{F}^{}\rangle$ is an eigenstate of 
$\mathcal{F}_z$ with an eigenvalue $m_{F}^{}=-F,\cdots,+F$. Assuming that the 
local Zeeman energy is minimized for $B_1'<0$, $B_0>0$, and 
$g_{\scriptscriptstyle F}<0$ we obtain in the $F=1$ case 
\begin{equation}
\label{berry} \bm{v}_s^{} =
\frac{\hbar}{m}\frac{1-\cos\vartheta'}{r'\sin\vartheta'}\hat{\bm{e}}_{\varphi'},
\end{equation}
where $(r',\varphi',\vartheta')$ refer to spherical coordinates in
the new coordinate system. Similar result has also been obtained
in Ref.~\cite{Savage:2003} for the hedgehog texture of Fig.~\ref{monopoles}(b). 

The superfluid velocity $\bm{v}_s^{}$
is equivalent to the vector potential of a magnetic monopole and
it has a Dirac string along the negative $z'$ axis if $B_0(t=0) >
0$. For spinor condensates the superfluid flow can be irrotational
and it is characterized by its vorticity $\bm{\Omega}_s^{} =
\nabla\times \bm{v}_s^{}$ which becomes
\begin{equation}
\label{omega} \bm{\Omega}_s^{} =
\frac{\hbar}{m}\frac{1}{r'^2}\,\hat{\bm{e}}_{r'}^{},
\end{equation}
indicating that vorticity corresponding to the imprinted monopole
defect is equivalent to the magnetic field of a magnetic monopole,
see Fig.~\ref{monopoles}. In particular, the topology of
$\bm{\Omega}_s^{}$ is unaffected by the scaling and translation,
and hence $\bm{\Omega}_s^{}$ remains equivalent to the hedgehog
texture shown in Fig.~\ref{monopoles}(b) also in the original
coordinate system. The Mermin-Ho relation for 
general spin-$F$~\cite{Ho:1996,Lamacraft:2008} can be computed from 
Eq.~\eqref{eigenstate} yielding
\begin{equation}
\label{mermin-ho}
\bm{\Omega}_S^{} = \frac{m_{F}^{}\hbar}{2m}\,\varepsilon_{ijk}^{}\,\hat{n}_i\,\nabla\hat{n}_j\times\nabla\hat{n}_k.
\end{equation}
In the adiabatic limit spin $\bm{\mathcal{S}}$ aligns with $\hat{\bm{n}}$ and 
the charge $\mathcal{Q}$ in Eq.~\eqref{charge} is thus directly proportional to the
flux of $\bm{\Omega}_s^{}$ through a surface $\Sigma$ enclosing
the defect. The flux of the monopole is supplied by the Dirac
string which is omitted form Eq.~(\ref{omega}). In case of $F=1$, the
monopole flux is $2h/m$, that is, two angular momentum quanta,
which implies that the vortex filament terminating at the monopole
must carry the same amount of vorticity. The situation is thus
similar to the Dirac monopole in superfluid 
$^{3}$He-A~\cite{Volovik:1976,Volovik:2003}. For a general
spin-$F$ BEC the Dirac string carries $2F$ quanta of angular
momentum.

Non-adiabatic effects arising from interactions between atoms,
kinetic energy, and the finite timescales in manipulating the
external magnetic fields can render the spin to deviate from the
direction of the local magnetic field. We take these effects into
account by solving the dynamics of the spinor order parameter from
the Gross-Pitaevskii mean-field equation~\cite{Ohmi:1998,Ho:1998}
\begin{equation}
\label{gpe} i\hbar\,\frac{\partial}{\partial t}\bm{\Psi}
=\big[\hat{h}_0 + c_0^{}|\bm{\Psi}|^2 +
c_2^{}(\bm{\Psi}^{\dagger}\bm{\mathcal{F}}\bm{\Psi})\cdot\bm{\mathcal{F}}\big]\bm{\Psi},
\end{equation}
where the interaction strengths $ c_0^{}$ and $ c_2^{}$ depend on
the scattering lengths in the different channels corresponding to
the total hyperfine spin of two scattering particles~\cite{Ho:1998}.
The single-particle operator $\hat{h}_0$ is given by
\begin{equation*}
\hat{h}_0=-\frac{\hbar^2}{2m}\nabla^2+ V_{\mathrm{opt}} +
g_{\scriptscriptstyle F}\mu_{\scriptscriptstyle
B}\bm{B}\cdot\bm{\mathcal{F}},
\end{equation*}
and in the simulation, we employ an external optical potential
$V_{\mathrm{opt}}=mr^2\omega_r^2/2$ and choose the parameters 
according to $^{87}$Rb which implies ferromagnetic interactions. 

In the simulation, energy time, and spatial variables are given in the units of 
$\hbar\omega_r^{}$, $1/\omega_r^{}$ and $a_r^{}=\sqrt{\hbar/m\omega_r^{}}$, 
respectively. For $\omega_r^{}=2\pi\times 250$, the dimensional values of the 
parameters are given by $B_1'=-0.05$ T/m and $B_0(t=0)=1$ \textmu T 
corresponding to $10^5$ atoms. We have also considered a different atom number 
and antiferromagnetic interactions~\footnote{For $^{87}$Rb we have obtained similar 
result for $N=5\times 10^4$ atoms. In the case of antiferromagnetic 
interactions, parameters were taken according to $^{23}$Na corresponding to 
$B_1'=-0.03$ T/m, $B_0(t=0)=1$ \textmu T, and $N=4\times 10^5$.}. The area 
considered in the simulation is $24 \times 24 \times 27$ in the units of 
$a_r^{}$ and the size of the computational grid varies from 
$141\times 141 \times 161$ to $175\times 175 \times 195$ points. In the 
simulation, we first calculate the ground state of
the system corresponding to the initial values of the magnetic
fields using the successive over relaxation (SOR) algorithm, and then 
propagate the initial state according to the time-dependent 
GP~equation~\eqref{gpe} using the split operator method combined with the 
Crank-Nicolson method. The time step used in the simulation is 
$10^{-4}/\omega_r^{}$.

\begin{figure}[h!]
\centering
\includegraphics[width=0.4\textwidth]{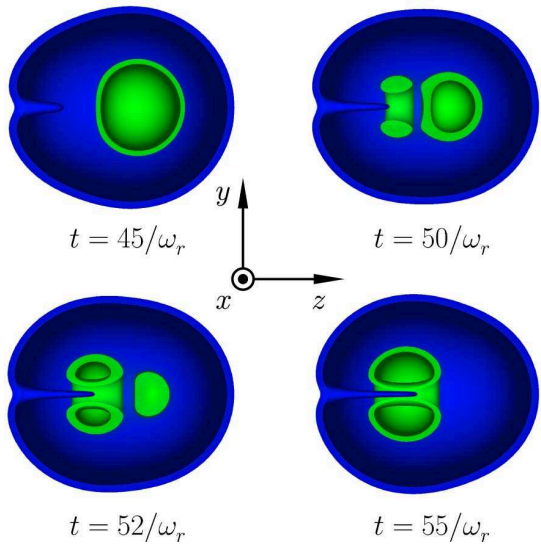}
\caption{\label{densities} Particle density at different stages 
  of monopole creation.  The monopole is created by ramping the bias field down
  in a period $t_0=50/\omega_r$. Blue (dark) color denotes densities from
  $\varrho = 3.8\times 10^{-5}$ $N/a_r^{3}$
  to $\varrho = 1.8\times 10^{-4}$ $N/a_r^{3}$
  and green (light) color densities from
  $\varrho = 1.4\times 10^{-3}$ $N/a_r^{3}$ to
  $\varrho =1.5\times 10^{-3}$ $N/a_r^{3}$. 
  The Dirac string is identified as the density depletion that propagates
  through the condensate.}
\end{figure}

Since the Dirac string carries two quanta of angular momentum it
is expected to be prone to splitting into two separate strings
each carrying one angular momentum quantum. To avoid this
scenario, we first imprint the monopole defect with the bias field
parallel to $z$ axis.
The particle density from this simulation is shown in
Fig.~\ref{densities}. The bias field is ramped linearly down to
zero in a period $t_0=50/\omega_r$ and then equally to negative
values. The spin texture $\bm{\mathcal{S}}$ takes the form
described in Fig.~\ref{monopoles}(a) and the corresponding
vorticity is shown in Fig.~\ref{vorticity}(a). Since the monopole
is brought in along the symmetry axis of the system, the Dirac
string remains intact. The created monopole appears to be
relatively stable: after the bias field was ramped down, the
monopole was allowed to evolve in time for a period
$5/\omega_r^{}$ in the presence of the two quadrupole fields
resulting in slow small-amplitude oscillations along the $z$ axis
but it remained otherwise intact. Due to nonadiabatic effects in
the creation process, the monopole defect in $\bm{\mathcal{S}}$
lags slightly behind the zero point of the magnetic field and the
Dirac string deviates from a pure $\delta$-function distribution.
We have also taken the parameters according to $^{23}$Na, i.e.,
corresponding to antiferromagnetic interactions, and obtained
essentially the same behavior as in the case of $^{87}$Rb.

\begin{figure}[h!]
\centering
\includegraphics[width=0.5\textwidth]{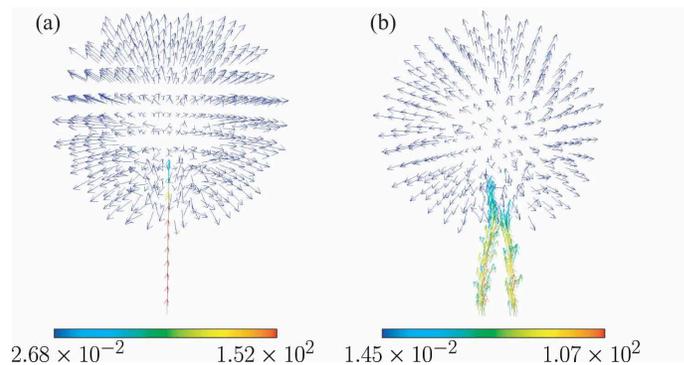}
\caption{\label{vorticity} Vorticity corresponding to the
  monopole. Vorticity
  $\bm{\Omega}_s^{}$ is computed numerically after the bias field is
  ramped from $B_0(0) = 6.7$ $\mu$T to $B_0(t_0) = 0$ and it is shown in
  the units of $h/{ma_r^{}}$. (a)
  $t_0=50/\omega_r^{}$ and the bias field is parallel to $z$ axis. (b)
  $t_0=40/\omega_r^{}$ and the bias field is parallel to the vector
  $3/2\,\hat{\bm{x}} + \hat{\bm{y}} + 14/3\,\hat{\bm{z}}$.
  Magnitude of the vorticity is denoted by color
  and the map is linear between the minimum and maximum values. 
  For clarity,
  only the relevant parts of the vector field $\bm{\Omega}_s^{}$ are
  shown.} 
\end{figure}

If the rotation symmetry with respect to $z$ axis is explicitly
broken by a bias field with $\hat{\bm{b}} \times \hat{\bm{z}} \neq
0$, we observe the Dirac string to split into two parts, see
Fig~\ref{vorticity}(b). The particle density is not depleted all
the way along the two strings suggesting that nonadiabatic effects
become more pronounced in this case. The spin density
$|\bm{\mathcal{S}}|$, on the contrary, exhibits a clear depletion
along the path of the two Dirac strings~\cite{EPAPS} and together
with the state-of-the-art experimental
methods~\cite{Higbie:2005,Sadler:2006} it should give an efficient
signature of the monopole. In the experiments, a typical
imperfection is a slight misalignment and slow drift between the
center of the optical trap and the symmetry axis of the magnetic
field. We model this by taking $\hat{\bm{b}} || \hat{\bm{z}}$ and
adding a small constant term $\tilde{B}_0\,\hat{\bm{x}}$ to the
magnetic field in Eq.~(\ref{mg-field}). This introduces an offset
equal to $7$\% of the effective radius of the condensate to the
path traced by the monopole with respect to $z$ axis. The offset
breaks the rotation symmetry of the system but the Dirac string of
this off-axis monopole remains undivided (data not shown)
suggesting that the instability of the string becomes visible only
for fairly large perturbations.

Since the monopole defect in the spin texture is topologically unstable, it 
can be removed by local surgery. Thus we address to the fate of the monopole 
after it has been created and all the magnetic fields pinning the monopole 
are turned off. In the case of superfluid
$^{3}$He,  Dirac monopoles are expected to drift to the boundary
of the vessel to form a surface
defect---boojum~\cite{Mermin:1977}. To analyze the decay of the
monopole in our case, we have carried out simulations in which
the external magnetic fields are switched off immediately or
ramped down with constant speed in a period $t_1$. For the finite
switch-off times, the decay of the monopole initiates already
while the external fields are being ramped down. The initial state
in the simulation corresponds to a monopole created by ramping the
bias field aligned with $z$ axis down in a period $t_0=50/\omega_r^{}$.

For ferromagnetic interactions, we have used three different
values $t_1=0$, $t_1=2.5/\omega_r^{}$ and $t_1=5.0/\omega_r^{}$.
In all three cases, the qualitative features are the same and they
are shown schematically in Fig.~\ref{decay} (see also~\cite{EPAPS}). 
For the monopole
defect in the spin texture, the monopole unwinds itself along the
Dirac string and results in a closed vortex ring, see
Fig.~\ref{decay}(a--d). In the course of unwinding, a cylindrical
domain wall separating the expanding core of the Dirac string from
the rest of the texture is formed. Depending on how fast the
external fields are turned off, the domain wall is either directly
pushed out of the condensate (slow turn-off) or contracted to
another vortex ring which eventually drifts out of the condensate
(rapid turn-off). In the simulations, the resulting vortex ring in spin
texture persists until the end of the simulation which in all
three cases spans a period $10/\omega_r^{}$. On the other hand,
the unwinding of the monopole in the vorticity $\bm{\Omega}_s$ is depicted in 
Fig.~\ref{decay}(e--h). During
the unwinding of the monopole, vorticity concentrated at the Dirac
string diffuses outwards and tends to relax towards more
uniformly distributed values.

\begin{figure}[h!] 
\centering 
\includegraphics[width=0.45\textwidth]{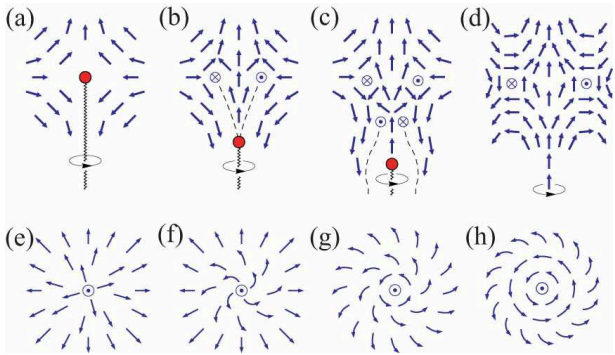} 
\caption{\label{decay}
Unwinding of the monopole defect in the spin texture 
$\bm{\mathcal{S}}$ (a)--(d) and in the vorticity $\bm{\Omega}_S^{}$ (e)--(h). 
Arrows pointing upwards (downwards) from the plane are denoted by 
$\bm{\odot}$ ($\bm{\otimes}$). Along the dashed line in (b) and (c), 
the spin points either upwards or downwards from the plane. The unwinding in 
$\bm{\Omega}_S^{}$ is shown in a plane perpendicular to $z$ axis and it is 
qualitatively independent of the $z$ coordinate.
High-resolution figures from the simulation corresponding to these schematic 
illustrations are available in~\cite{EPAPS}.}  
\end{figure}  

For antiferromagnetic interactions we have carried out a
simulation with magnetic fields ramped down linearly in a period
$t_1=2.5/\omega_r^{}$. The results agree qualitatively with the
ferromagnetic case and the end configuration is again a vortex
ring in the spin texture. Another vortex ring is generated at the
boundary of the condensate and it persists until the end of the
simulation. The length of the simulation was $6/\omega_r^{}$ in
this case. It should be noted that simulations concerning
the unwinding of the monopole span a relatively short period and
thus it remains a question for the future research to study if the
evolution of the created monopoles differs qualitatively between
ferromagnetic and antiferromagnetic interactions. From the global
symmetries of the order parameter in the absence of external
fields~\cite{Ho:1998,zhou}, one could expect different behavior by
purely topological reasoning~\cite{Chuang:1991}. Furthermore, we have not 
included the possible magnetic dipole-dipole interaction which may change the 
dynamics of the monopole unwinding. In particular, the spin 
texture considered here resembles the co-called two-z-flare texture that was 
found to be stabilized by the dipole-dipole interaction~\cite{Takahashi:2007}.

In conclusion, we have introduced and modelled a robust method to
create Dirac monopoles in spinor Bose-Einstein condensates. Our
proposal is directly realizable in the present-day experiments and
recent developments in the non-destructive imaging of the
magnetization of spinor BECs~\cite{Higbie:2005,Sadler:2006} pave
way for detecting the exciting dynamics of these monopoles such as
the unwinding of the monopole. 

Authors acknowledge Jenny and Antti Wihuri Foundation, Emil Aaltonen 
Foundation,and the Academy of Finland for financial support, and the 
Center for Scientific Computing Finland (CSC) for computing resources. 
We thank M.~Krusius, G.~Volovik, W.~Phillips, and P.~Clad\'{e} for 
discussions, and J.~Huhtam\"{a}ki for help in numerical calculations.

\bibliography{manu}

\setcounter{figure}{0}

\begin{titlepage}

$ $

\vspace{3cm}

\centering

{\bf \LARGE Supplementary Figures -- Creation of Dirac monopoles in spinor Bose-Einstein condensates}

\vspace{1cm}

{\Large Ville Pietil\"a$^{1,2}$ and Mikko M\"ott\"onen$^{1,2,3}$}

\vspace{5mm}

$^{1}$Department of Applied Physics/COMP, Helsinki
University of Technology P.~O.~Box 5100, FI-02015 TKK, Finland 

\vspace{2mm}

$^{2}$Australian Research Council, Centre of Excellence for Quantum Computer Technology,
The University of New South Wales, Sydney 2052, Australia 

\vspace{2mm}

$^{3}$Low Temperature
Laboratory, Helsinki University of Technology, P.~O.~Box 3500,
FI-02015 TKK, Finland

\end{titlepage}

\pagebreak

\pagestyle{empty}

\begin{figure*}[t!]
\centering
\includegraphics[width=0.8\textwidth]{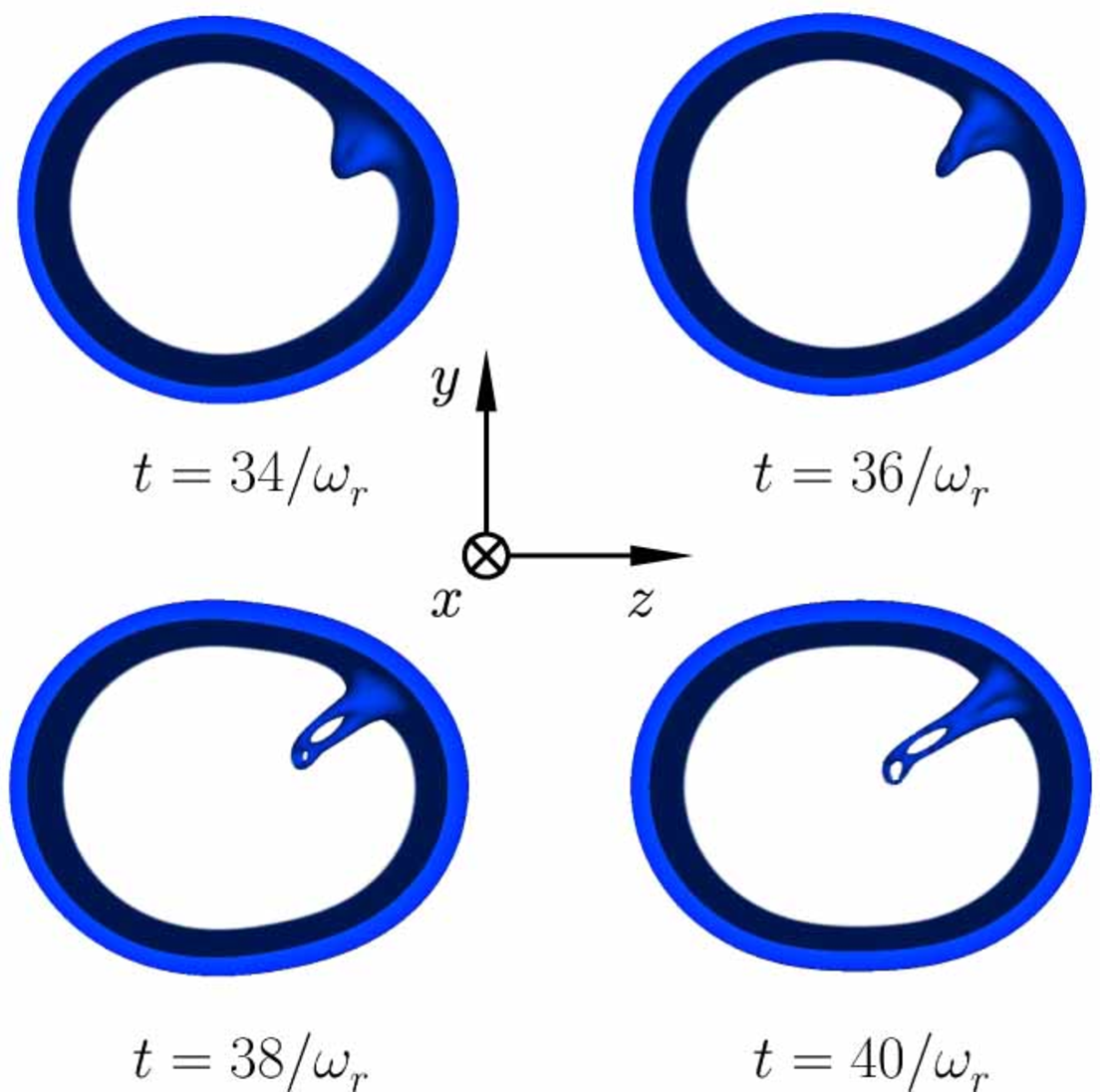}
\caption{\label{spin_densities} {\bf Spin density  of the
condensate}. Splitting of the Dirac string is manifested in the spin density
$|\bm{\mathcal{S}}|$ which is depleted along the two strings.
The figure corresponds to spin densities from
$|\bm{\mathcal{S}}| = 1.5\times 10^{-4}$ $N/a_r^{3}$ to
$|\bm{\mathcal{S}}| = 4.0\times 10^{-4}$ $N/a_r^{3}$. Here $x$ axis points
downwards from the plane.}
\end{figure*}

\pagebreak

\begin{figure*}[t!]
\centering
\includegraphics[width=1.0\textwidth]{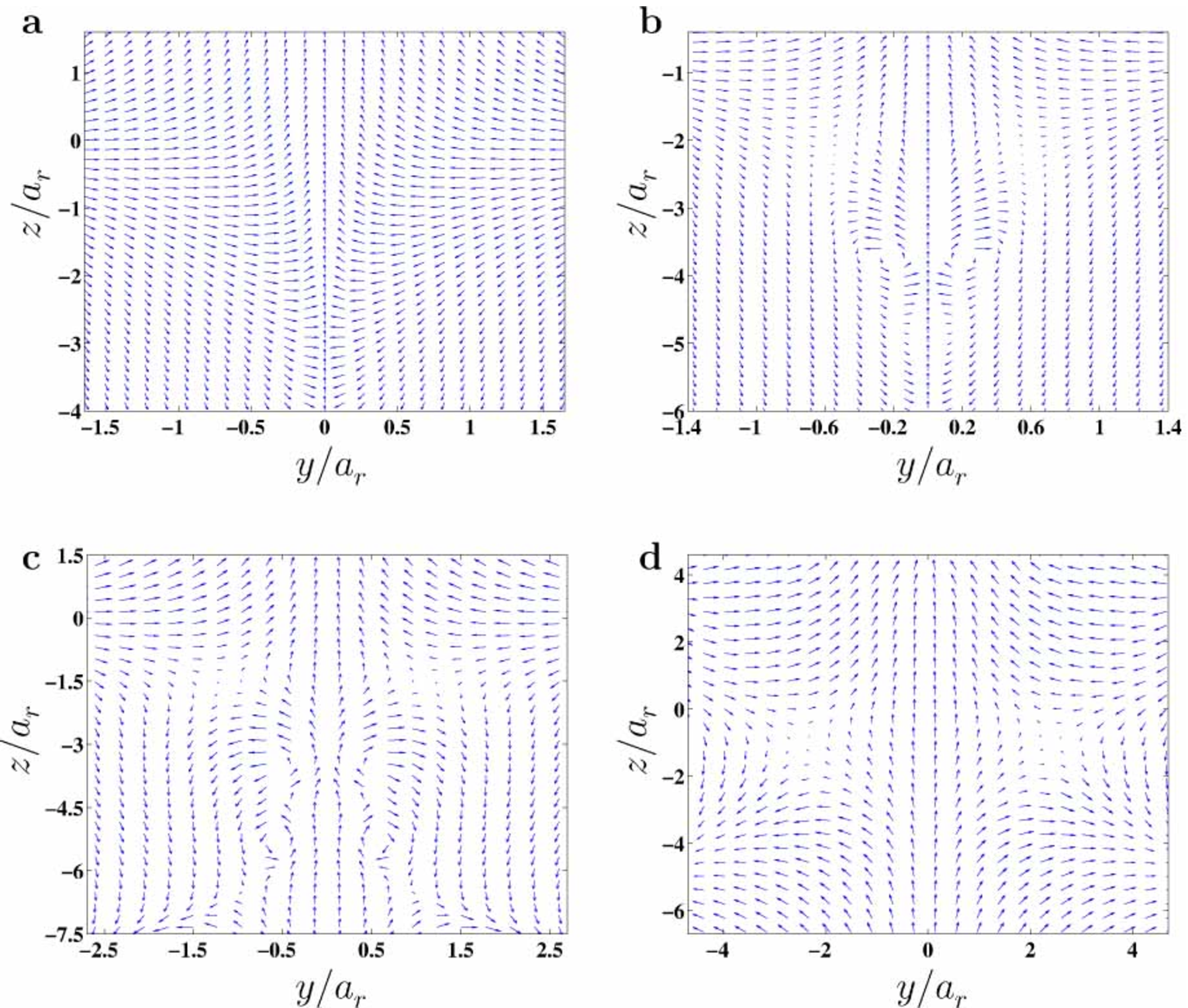}
\caption{\label{spin} {\bf Unwinding of the monopole defect in the spin
texture}. The 2D plane corresponds to $x=0$. In the figure, $\hat{s}_y$ and $\hat{s}_z$
components of the unit spin vector
$\hat{\bm{s}}=\bm{\mathcal{S}}/|\bm{\mathcal{S}}|$ are shown.
The spatial scale in different panels changes since the radius of the vortex
ring changes during its formation. Time corresponding to each panel is
$50.5/\omega_r^{}$ (a), $52/\omega_r^{}$ (b), $53/\omega_r^{}$
(c), and $60/\omega_r^{}$ (d).}
\end{figure*}

\pagebreak

\begin{figure*}[t!]
\centering
\includegraphics[width=1.0\textwidth]{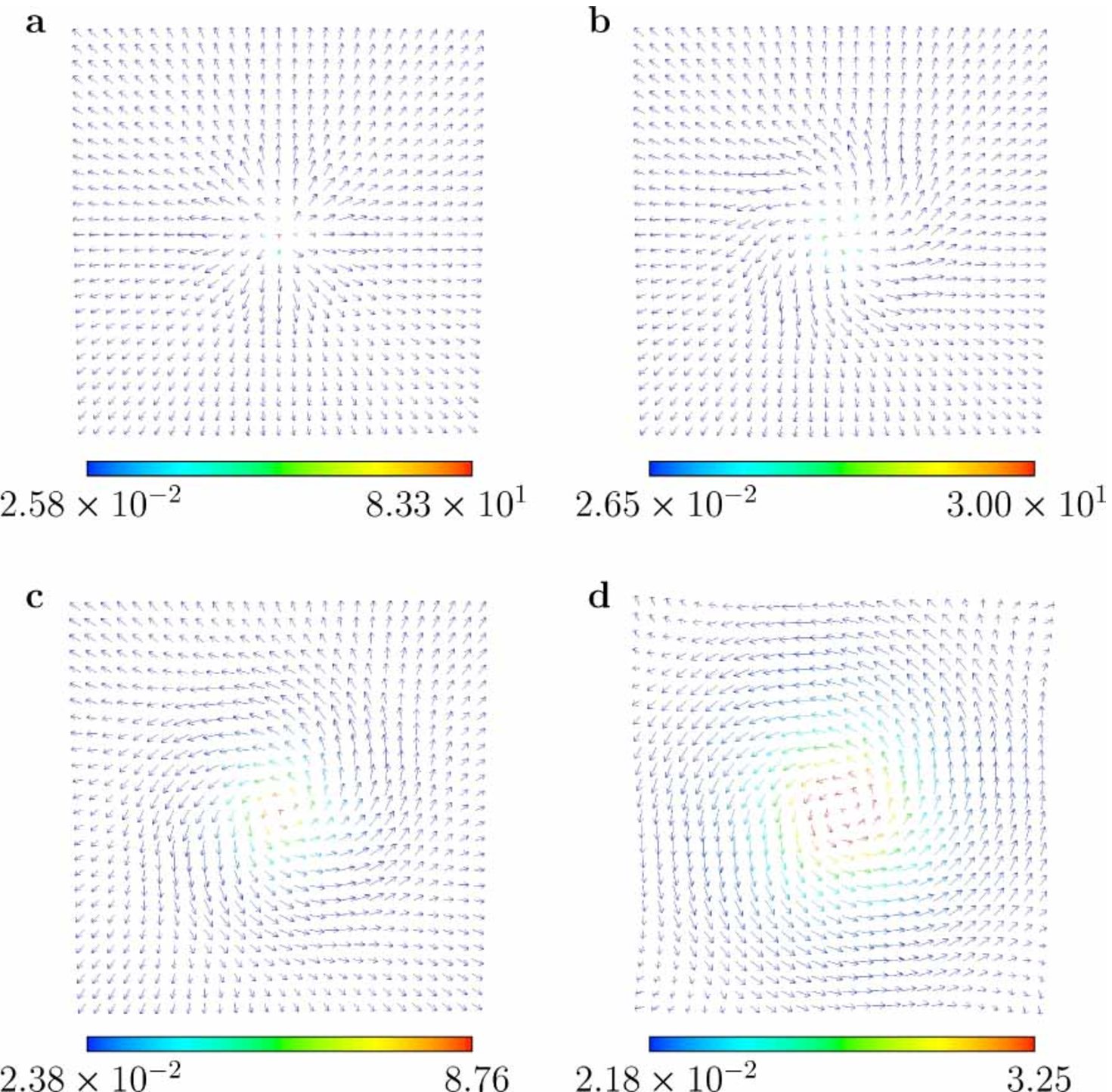}
\caption{\label{gauge} {\bf Unwinding of the Dirac monopole in vorticity}.
The 2D plane corresponds to $z=-2.8$ $a_r^{}$. Vorticity is given in the units
of $h/{ma_r^{}}$. The time instant corresponding to each panel is
$50.5/\omega_r^{}$ (a), $51.5/\omega_r^{}$ (b), $52.5/\omega_r^{}$
(c), and $53.5/\omega_r^{}$ (d). At the center of the defect, the
vector field is always parallel to the $z$ axis pointing upwards from the
plane. In the figure $x$ and $y$ axes correspond to horizontal and vertical
axes, respectively. The field of view is $15a_r^{}\times 15a_r^{}$.}
\end{figure*}

\end{document}